\documentstyle[aps,multicol,prl,psfig]{revtex}

\draft

\begin{document}

\newcommand{\be}{\begin{eqnarray}}
\newcommand{\ee}{\end{eqnarray}}

\title{Kinematic reduction of reaction-diffusion fronts
with multiplicative noise: \\
Derivation of stochastic sharp-interface equations}

\author{A. Rocco$^{1,2}$, L. Ram\'{\i}rez-Piscina$^{3}$ and 
J. Casademunt$^{2}$} 
\address{$^{1}$CWI, Postbus 94079, 1090 GB Amsterdam, The Netherlands}
\address{$^{2}$Departament d'Estructura i Constituents de la Mat\`{e}ria, 
Facultat de F\'{\i}sica, Universitat de Barcelona, Av. Diagonal 647,
E-08028 Barcelona, Spain}
\address{$^{3}$Departament de F\'{\i}sica Aplicada, Universitat 
Polit\`{e}cnica de Catalunya, Avenida Dr. Gregorio Mara\~n\'on 50,
E-08028 Barcelona, Spain}

\date{\today}

\maketitle

\begin{abstract}
We study the dynamics of generic reaction-diffusion 
fronts, including pulses and chemical waves, 
in the presence 
of multiplicative noise.
We discuss the connection between the reaction-diffusion Langevin-like 
field equations 
and the kinematic (eikonal) description in terms of a 
stochastic moving-boundary or sharp-interface approximation.
We find that the effective noise is additive and we relate its  
strength to the noise parameters in the original field equations, to 
first order in noise strength, but including a partial resummation to
all orders which captures the singular dependence on the microscopic 
cutoff associated to the spatial correlation of the noise. 
This dependence is essential for a quantitative and qualitative
understanding of fluctuating fronts, affecting both scaling properties 
and nonuniversal quantities. Our results
predict phenomena such as  
the shift of the transition point between the pushed and pulled 
regimes of front propagation,
in terms of the noise parameters, and the corresponding transition
to a non-KPZ universality class.
We assess the 
quantitative validity of the results in several examples 
including equilibrium fluctuations, 
kinetic roughening,
and the noise-induced pushed-pulled transition, which is predicted and 
observed for the 
first time.
The analytical predictions 
are successfully tested against rigorous results and show excellent
agreement with 
numerical simulations of reaction-diffusion field equations with multiplicative 
noise. 
\end{abstract}


\section{Introduction}
The dynamics of localized solutions in the form of fronts or  
pulses in reaction-diffusion systems has received a great deal of attention 
for a long time in the context of nonequilibrium extended systems 
\cite{Cross}. Examples of fronts formed by stable regions  
invading unstable or metastable ones 
are found in a large variety of physical, chemical or biological systems,
and have been studied in great detail \cite{vanSaarloos,Collet90,Ebert00a}. 
In the context 
of excitable media and chemical waves, extended pulses do also exhibit a rich 
phenomenology\cite{Kapral93,Mikhailov95,tyson,Meron,Mikhailov94}. 
As opposed to fronts, excitation waves or pulses are such 
that the region behind them 
eventually returns to the same linearly stable state that is 
ahead.  
In this case, in dimensions higher than one there may be open ends of the 
pulse region which give rise to spiral (2d) or scroll (3d) waves.
There has been much interest in the study of such objects from the fundamental 
point of view of pattern forming dynamics, but also because of their potential
applications in biological systems, such as in cardiac tissue \cite{heart}.
In this paper we will not consider the case of 
open ends, so unless otherwise specified, we will refer indistinctly to 
fronts and pulses under the common term of 'fronts'.

One aspect which has received increasing interest in recent years 
has been the effect of fluctuations both of internal and external origin 
in the dynamics and in the 
roughening properties of fronts 
\cite{Riordan95,Armero96,Armero98,Kessler98,Rocco00,Tripathy00,Tripathy01,Rocco01,vulpiani}. 
More recently the effect of noise in pattern 
forming dynamics of chemical waves has been 
fostered by the development of the experimental capability to introduce 
external spatio-temporal 
noise in a controlled way in different photosensitive 
nonlinear chemical reactions, through the optical 
projection of computer-designed spatio-temporal fluctuations in the  
local illumination conditions, acting as a stochastic 
control parameter 
\cite{Sendina-Nadal97,Sendina-Nadal98,Sendina-Nadal00,Showalter98}.
From a theoretical point of view, a common starting point to study 
fluctuations 
is in terms of master equations defining evolution of reacting and diffusing 
particles in a lattice\cite{Riordan95}. 
The connection of this microscopic level of description 
to the mesoscopic one in terms of Langevin field 
equations has proven a rather
subtle issue, in connection with the distinction of the so-called pushed vs 
pulled fronts \cite{Ebert00a}. Only recently a complete understanding
of the instrinsically 
different nature of these two types of fronts, and the corresponding 
consequences
concerning the presence of cutoffs \cite{Brunet97,Kessler98b,Panja01,Moro01}, 
and the effects of fluctuations
has started 
to emerge \cite{Armero96,Armero98,Rocco00,Tripathy00,Tripathy01}. 
It has been shown, for instance, that pulled fronts define a new
universality class of kinetic roughnening different 
from the Kardar-Parisi-Zhang 
universality class \cite{Tripathy00,Tripathy01}.
On the other hand, it has been shown that intrinsic noise 
at the microscopic level may induce a morphological instability at the 
macroscopic level of description\cite{Kessler98}. 
In this paper we will be mostly concerned
with the macroscopic description of pushed fronts with fluctuations, but 
also on how this description incorporates the transition to pulled fronts 
induced by the noise itself.

In the absence of noise, and in the appropriate limit, 
the description of fronts and pulses defined by reaction-diffusion 
field equations can be reduced by means of a moving boundary approximation 
to a kinematic description in terms of much simpler local equations 
\cite{Mikhailov94,hakim}. This procedure, which is mathematically well grounded
in the framework of the so-called inertial manifold reduction, 
can be carried out systematically, for instance, through asymptotic matching 
techniques using the front thickness as a small parameter defining a singular 
perturbation problem \cite{Ebert00b}. This 
is commonly used to relate macroscopic
interface equations to order-parameter or phase-field descriptions in 
many problems involving interface dynamics, such as in solidification
\cite{Wang93,Karma96}, 
viscous fingering \cite{folch} etc. 
In the case of chemical pulses in excitable media
in the limit of weak excitability, this leads to a local equation 
where the normal velocity of the pulse is a constant plus a correction 
proportional to curvature \cite{Mikhailov94}. 
This is often called the eikonal equation.
In the case of open ends, a similar local equation can be derived for 
the motion of the endpoint of the pulse \cite{Mikhailov94,hakim}.
For pushed fronts it can be shown that for smooth, long wavelength deformations
of the front, the separation of time scales between the soft deformation modes and
the internal degrees of freedom of the fields leads naturally to the same
eikonal equation. For pulled fronts, however, this separation of time scales
does not exist, the relaxation being algebraic instead of exponential, 
and a local moving-boundary approximation is not justified \cite{Ebert00b}.

The kinematic description in terms of eikonal-like equations is a very 
useful approximation from both a theoretical and a practical point of 
view. In the context of the study of universality of fluctuation 
properties \cite{stanley,Krug97} 
for instance, it leads naturally to the relevant effective universal 
description of a broad class of systems in terms of the KPZ equation 
\cite{Kardar86}.
It is also very useful for numerical simulation purposes, since it 
avoids resolving the fine structure of the front and the bulk degrees 
of freedom, which become 
irrelevant, 
dealing only with the kinematic degrees of freedom of an object
of reduced dimension. 

When noise is present in the original field equations, however, the 
situation is not so clear. Stochastic eikonal equations have 
proven useful 
in a phenomenological description of the dynamics of
pulses and spiral waves in photosensitive chemical systems with external 
noise imposed on the illumination conditions 
\cite{Sendina-Nadal98,Sendina-Nadal00}. 
Such description, however, relied on some fitting parameters and some 
uncontrolled hypothesis on the way the noise must enter 
the kinematic 
equations. These results, together with the fact that the statistical 
properties of the noise present in those experiments are fully controlled,
clearly call for a more 'microscopic' derivation of 
stochastic kinematic equations corresponding to Langevin reaction-diffusion 
field equations with multiplicative noise, with no free parameters. 
The connection between bulk and interface fluctuations has been worked 
out so far only for equilibrium fluctuations in coarse-grained, 
Ginzburg-Landau-like  
equations\cite{Hohenberg,Safran94,karmaprl,rappel}. 
In such cases, the sharp-interface 
limit can be taken at the level of the free-energy, and the existence of a 
fluctuation-dissipation theorem then yields the proper way to incorporate 
thermal fluctuations into the effective interface equations. However, in many 
nonequilibrium systems, for instance in the context of reaction-diffucion problems, 
a free-energy or generically a Liapunov functional may 
not exist and no fluctuation-dissipation relation may be invoked for external 
noise. In such cases the connection between the bulk description and the effective 
interface description in presence of fluctuations must be worked out at the 
level of the dynamical equations.

The purpose of this paper is to 
address this point by deriving stochastic 
eikonal equations, including the complete specification of the noise statistics 
in connection to that of the noise in the mesoscopic field equations.
In particular we shall focus on the singular dependence on the spatial 
cutoff when noise is multiplicative, and its importance in the quantitative
description of the statistical properties of the front fluctuations.
The predictions will be tested against numerical simulations of 
reaction-diffusion equations, and also in cases where exact 
results are available concerning the spectrum of interface fluctuations. 
We will also see that the stochastic eikonal 
equation derived is consistent with 
the scenario of the pushed-pulled transition, and that, changing the 
spatial cutoff or the noise intensity may have effects such drastic 
as changing the universality
class of kinetic roughening of the front 
through that transition.

Although our obtention of the stochastic eikonal equation here presented
is not a first-principles rigorous derivation, we will provide sufficient
evidence to conclude that the result is the correct one to lowest order
in noise intensity, including the singular dependence on the noise 
correlation cutoff (which involves a partial resumation to all orders), 
within the long time and length scales limits  
which are inherent to the kinematic descrition itself.
However,
it does not apply to situations in which the front dynamics is 
nonlocal, such as solidification fronts or viscous fingers, where a
different type of formulation is appropriate even in the deterministic case
\cite{Wang93,Karma96,folch,cinca}.

\section{Kinematic reduction for generic 
reaction-diffusion systems}

Let us consider a vectorial field ${\boldmath{\mbox{$\phi$}}}({\bf
x},t)$ with $N$ components 
${\boldmath{\mbox{$\phi$}}}({\bf x}) \equiv \phi_1({\bf x}),...,
\phi_N({\bf x})$ in a $d-$dimensional space with 
${\bf x} \equiv  x_1,...,x_d$, which 
obeys a reaction-diffusion equation with 
multiplicative noise of the form:
\be
\frac{\partial \boldmath{\mbox{$\phi$}}}{\partial t} = 
\hat{D} \nabla^2 \boldmath{\mbox{$\phi$}} + 
{\bf f}(\boldmath{\mbox{$\phi$}}) 
+ \varepsilon^{1/2} {\bf g}(\boldmath{\mbox{$\phi$}}) 
\eta({\bf x},t), \label{multfront}
\ee 
where $\eta({\bf x},t)$ is a gaussian noise with zero mean and 
correlation given by
\be
\langle \eta({\bf x},t) \eta({\bf x}^{\prime},t^{\prime}) \rangle = 
2 \lambda^{-d} C(|{\bf x}-{\bf x}^{\prime}|/\lambda) \delta(t-t^{\prime}).
\label{noisecorr}
\ee

We take a one-component noise for simplicity, as the natural case when it 
originates in fluctuations of a single control parameter. The generalization 
of the formalism to multicomponent noise is straightforward.
Notice the asymmetry with which we treat the spatial and temporal
correlator of the noise. As we will see, this reflects a nontrivial issue related 
to the intrinsically different nature of the
white noise limit in space as opposed to time. We have taken in Eq.
(\ref{noisecorr}) the noise as delta-correlated in time.
This temporal white noise limit is well behaved, once a prescription
for the multiplicative noise term in
Eq. (\ref{multfront}) has been chosen. 
For external fluctuations the physically relevant prescription
is to consider the white noise as the limit of some properly
defined correlated noise. This corresponds to the well-known Stratonovich
prescription \cite{Gardiner}. 
On the other hand, the spatial noise must always be defined as 
colored, its white (uncorrelated) limit being intrinsically ill-defined. Hence
the notation with the function $\lambda^{-d}C(r/\lambda)$ in the correlator,
meaning a general correlation function, dependent on some correlation length
$\lambda$, which in the limit $\lambda \rightarrow 0$ is such that 
$\lambda^{-d}C(r/\lambda) \rightarrow \delta(r)$.
The fact that the spatial continuum limit 
$\lambda \rightarrow 0$ does not exist in the Stratonovich interpretation
is thus reflected in the fact that, even if $\lambda$ is much smaller than 
any other length scale in the problem, the existence of such microscopic cutoff 
always shows up in the quantitative description of the large scale behavior and
cannot be reabsorbed in a redefinition of parameters.
That is, the noise cannot be considered as effectively white in space, 
regardless of how small the noise correlation length is.

The multiplicative noise term present in Eq. (\ref{multfront}) has an
average value different from zero. Applying Novikov theorem
\cite{Novikov64}, we get 
\be
\varepsilon^{1/2} \langle {\bf g}(\boldmath{\mbox{$\phi$}}) \eta({\bf x},t)
\rangle = \varepsilon \lambda^{-d}
C(0) \langle {\bf G(\boldmath{\mbox{$\phi$}})} \rangle 
\ee
where
\be
G_i(\boldmath{\mbox{$\phi$}}) \equiv \sum_j \frac{\partial
g_i(\boldmath{\mbox{$\phi$}})}{\partial
\phi_j} g_j(\boldmath{\mbox{$\phi$}}). \label{product}
\ee
This suggests to separate the average contribution from the
multiplicative noise term and rewrite Eq. (\ref{multfront}) in terms of a
renormalized potential and a zero average noise,
\be
\frac{\partial \boldmath{\mbox{$\phi$}}}{\partial t} = 
\hat{D} \nabla^2 \boldmath{\mbox{$\phi$}} + 
{\bf h}(\boldmath{\mbox{$\phi$}}) + \varepsilon^{1/2}
\boldmath{\mbox{$\Omega$}}(\boldmath{\mbox{$\phi$}}, 
{\bf x},t), \label{main}
\ee
where
\be
{\bf h}(\boldmath{\mbox{$\phi$}}) =
{\bf f}(\boldmath{\mbox{$\phi$}}) + \varepsilon \lambda^{-d}
C(0) {\bf G}(\boldmath{\mbox{$\phi$}}) \label{functh}
\ee
and
\be
\boldmath{\mbox{$\Omega$}}(\boldmath{\mbox{$\phi$}}, {\bf x},t) = 
{\bf G}(\boldmath{\mbox{$\phi$}})\eta({\bf x},t) 
- \varepsilon^{\frac{1}{2}} \lambda^{-d} C(0) {\bf G}(\boldmath{\mbox{$\phi$}}).
\ee 
Here the new noise $\boldmath{\mbox{$\Omega$}}$ has zero average.
Notice that this decomposition corresponds to transform Eq. (\ref{multfront}) into
its equivalent It\^o stochastic equation in the white noise limit.
Accordingly, the stochastic term $\Omega$ reduces to
\be 
\lim_{\lambda \rightarrow 0} \boldmath{\mbox{$\Omega$}}(\boldmath{\mbox{$\phi$}}, 
{\bf x},t) = {\bf G}(\boldmath{\mbox{$\phi$}})\eta_I({\bf x},t)
\ee
where $\eta_I({\bf x},t)$ is now a white noise in the It\^o interpretation. 
The deterministic term ${\bf h}$ includes thus noise effects through the so-called
Stratonovich term that has been added to ${\bf f}$. These noise effects on the
deterministic part of Eq. (\ref{main}) depend on a 'dressed' noise intensity
$\varepsilon_{\lambda}$ which contains the singular dependence on the  
the spatial cutoff in the form  
of the 'bare' $\varepsilon$ as $\varepsilon_{\lambda} \equiv \varepsilon C(0) 
\lambda^{-d}$. 
The important point is that the spatial white noise limit in 
the continuum equations is mathematically well-defined for an It\^o noise. This 
has been proven rigorously for relatively broad classes of equations (see for
instance Refs.\cite{Gyongy98,Nualart99}). On the contrary, this cannot be the case 
for a Stratonovich noise, as it is obvious from the singular 
dependence on $\lambda$. The practical implications of these fact are that, 
while the singular contribution of $\lambda$ in the Stratonovich term must be kept 
explicitly, the dependence on 
$\lambda$ contained in 
$\boldmath{\mbox{$\Omega$}}$ will indeed be weak (nonsingular), 
and indeed  negligible if $ \lambda$ is much smaller than the other length 
scales in the problem, in particular the front thickness. Therefore, the 
splitting  
of the Stratonovich noise by means of a deterministic term plus an It\^o noise 
has the virtue to isolate the singular dependence on the microscopic cutoff.
This will also define a useful partial resummation of orders 
in $\varepsilon_{\lambda}$. After this decomposition, the theory will thus 
be correct to first order in $\varepsilon$, to all orders in $\varepsilon/\lambda$ 
and to order $\lambda^0$ for the regular dependence of 
$\boldmath{\mbox{$\Omega$}}$ in the noise correlation length.

Eq. (\ref{main}) will be our starting point. We are interested in representing the
dynamics of a front obeying Eq. (\ref{main}) as 
the evolution of a $d-1$ surface.
In this effective dynamics we assume the details of the front structure (at scales
of order of or smaller than the front thickness) to be unimportant.
In what follows we will consider the evolution of a 1-dimensional front embedded
in a 2-dimensional system. In the procedure we are going to apply we will write
the evolution equation (\ref{main}) in the curvilinear 
coordinates defined by the
1d front in the sharp-interface limit, and obtain the evolution equation for this
front as a solvability problem
with the basic assumption that curvature and noise be small perturbations.

Before proceeding with the formal derivation, let us first point out some 
subtleties related to the stochastic case as opposed to the deterministic 
one. In the latter, it is customary to define a curvilinear coordinate 
system $(s,r)$ in which $r=0$ stands for the curve representing the front
position, which can be associated for instance to a level curve of the appropriate
field. The scheme assumes that the front thickness is small compared to the 
radius of curvature, and that the relaxation of the internal degrees of 
freedom of the front is fast compared to the time scale of the long wavelength 
deformations of the front.
When noise is present in the field equation, the appropriate curvilinear 
coordinate system cannot be associated to level curves, which are very rough 
at length scales smaller than the front thickness. On the other hand, at larger 
scales, which are the ones we are interested in, a coarse 
grained description makes perfect sense. This is actually implicit in the 
very idea of the stochastic eikonal equation. One can think of different 
schemes to explicitly define such coarse-grained description, all of them 
equivalent. However, since the rest of the derivation cannot be carried 
out explicitly in full rigor in any of those, and since the result is 
expected to be 
independent of the details of the definition of the coarse-graning, we will
proceed more or less heuristically. 
In essence this is a reformulationn of the approach introduced for the  
derivation of the diffusive wandering of fronts in one dimension 
discussed in Ref.\cite{Armero98}. There, the basic idea was that only the 
low-frequency components of the noise are responsible for the front wandering 
so the high-frequency components can be implicitly integrated out. The effect
of the high-frequency components of the noise is thus to renormalize the 
mean front profile. As a consequence, they renormalize the front velocity, and 
also the diffusion coefficient. 
More precisely, this means in our case that the 
high-frequency renormalization is carried out by the Stratonovich term in the 
function ${\bf h}$. Once this term is explictly extracted, the remainder is an 
It\^o multiplicative noise. Then the high-frequency components 
of this term are irrelevant and can be averaged out, while the low-frequency 
components will be responsible for the roughnening of the front at scales
larger than the coarse-graining length. 
In the case of 1d fronts, the resulting 
diffusion coefficient for the front wandering has been rederived rigorously 
in Ref.\cite{Rocco00}. Unfortunately, the identification of the collective 
variable in 1d cannot be immediately generalized to higher dimensions, so 
we must still rely on a less precise formulation and check the consistency with 
rigorous results and numerical simulations \it a posteriori\rm.

After the above considerations, we can make the following theoretical 
construction. We assume we have solved the 
field equations with noise without any approximation. We now 
coarse grain the fields with some local 
average procedure, both in time and space, and use the coarse-grained fields to 
define a curvilinear 
coordinate system based, for instance, in terms of level curves at any time. 
At short length and time scales 
this coordinate system is smooth and, in principle,
we could write the full field equation 
(still with the bare fields) in these curvilinear coordinates.
In the absence of noise, 
an expansion in the front thickness 
would unambiguosly yield the terms which are dominant in the range where the 
eikonal equation is devised for, namely for sufficiently long length scales 
(small curvatures) and long time scales.
Terms such as second derivatives on $s$ and the time derivative would
automatically drop out of the description.
In the presence of noise stochastic case, this is not so automatic unless 
the fields themselves are coarse-grained so that they are also
sufficiently smooth in space and time. 
We assume that, for the coarse-grained fields,
the order of the different terms in the front thickness will be the same than 
for the deterministic case, when such expansion makes sense (excluding pulled 
fronts, for instance). We claim that such assumption is the one implicit in the 
very idea of the existence of a stochastic kinematic formulation of the front
dynamics.
Then, for the coarse-grained fields, Eq. (\ref{main}) is expected to reduce,
in analogy to the deterministic case, to 
\be
\hat{D} \frac{\partial^2 \boldmath{\mbox{$\phi$}}}{\partial r^2} + 
\hat{D} \kappa(s,t) \frac{\partial\boldmath{\mbox{$\phi$}}}{\partial r} 
+ {\bf h}(\boldmath{\mbox{$\phi$}}) +
v_n(s,t) \frac{\partial \boldmath{\mbox{$\phi$}}} {\partial r}
&+& 
\varepsilon^{1/2} \boldmath{\mbox{$\Omega$}}(\boldmath{\mbox{$\phi$}}; r,s;t) 
= 0,
\label{wtho} 
\ee
where $\kappa$ is the local curvature and $v_n$ the normal velocity of the front. 
This normal velocity provides the evolution of the curvilinear
coordinates in which
Eq. (\ref{main})  takes the form of Eq. (\ref{wtho}), and is the
fundamental quantity we are interested
in. The noise term in Eq. (\ref{wtho}) must also be considered as 
coarse-grained, with the high-frequency high-wavenumber 
components integrated out. 

At this point of the derivation it is useful to consider the 1d 
problem which 
corresponds to the zeroth order of the eikonal description. This is defined
by neglecting the curvature and the fluctuating term in Eq. (\ref{wtho}): \be
0 = \hat{D} \frac{\partial^2 \boldmath{\mbox{$\phi_0$}}}{\partial r^2} 
+ \bar{v}(\varepsilon_\lambda) \frac{\partial \boldmath{\mbox{$\phi_0$}}}{\partial r} +
{\bf h}(\boldmath{\mbox{$\phi_0$}}).  \label{fi}
\ee
This is the eigenvalue problem that gives the renormalized velocity $\bar{v}$
in the 1d problem as obtained in Refs. \cite{Armero96,Armero98}. Note that this
equation does contain noise effects through the high-frequency renormalization 
provided by the Stratonovich term.
In fact, the effective velocity $\bar{v}(\varepsilon_\lambda)$ resulting 
from Eq. (\ref{fi}) has contributions from all orders in the dressed noise 
intensity
$\varepsilon_\lambda$. An explicit first order in $\varepsilon_\lambda$
approximation will be given in the Appendix.
In the problem defined by Eq. (\ref{wtho}) curvature and  
fluctuations will be taken as small perturbations.
Hence we assume for the field $\boldmath{\mbox{$\phi$}}$ and the
velocity $v_n$ expansions of the form
\be
\boldmath{\mbox{$\phi$}}(r,s,t) = \boldmath{\mbox{$\phi_0$}}(r) 
+ \delta\boldmath{\mbox{$\phi$}}(r,s,t),
\label{pe}
\ee
\be
v_n(s,t) = \bar{v}(\varepsilon_\lambda)
+ \beta(\varepsilon_\lambda) \kappa(s,t)
+ \delta v(s,t), \label{scurv}
\ee
where $\boldmath{\mbox{$\phi_0$}}(r)$ and $\bar{v}(\varepsilon_\lambda)$ 
are the solution of the
1d problem of Eq. (\ref{fi}). The term $\beta \kappa(s,t)$ is a
curvature correction and $\delta v(s,t)$ describes fluctuations.
Linearizing in both perturbations $\delta{\boldmath{\mbox{$\phi$}}}(r,s)$ 
then verifies
\be
0 = \hat{\Gamma} \delta \boldmath{\mbox{$\phi$}} 
+ (\beta \kappa + \delta v) \frac{\partial \boldmath{\mbox{$\phi_0$}}}{\partial r}
+ \hat{D} \kappa \frac{\partial \boldmath{\mbox{$\phi_0$}}}{\partial r} 
+ \varepsilon^{1/2} \boldmath{\mbox{$\Omega$}}(\boldmath{\mbox{$\phi_0$}}; r,s;t), 
\label{deltafi}
\ee
where
\be
\hat{\Gamma} = \hat{D} \frac{\partial^2}{\partial r^2} 
+ \bar{v}(\varepsilon_\lambda)
\frac{\partial}{\partial r} + \left.\frac{\partial {\bf h}}
{\partial \boldmath{\mbox{$\phi$}}} \right|_{ \boldmath{\mbox{$\phi$}} 
= \boldmath{\mbox{$\phi_0$}}}. 
\label{gamma}
\ee

Taking the derivative of Eq. (\ref{fi}) with respect to $r$, it is a
simple matter to prove that 
\be
{\bf u_0} = \frac{\partial \boldmath{\mbox{$\phi_0$}}}{\partial r}
\ee 
is the right eigenvector of $\hat{\Gamma}$ with zero eigenvalue. 
Due to the non-hermiticity of $\hat{\Gamma}$, finding
an analytic expression for the left eigenvector, ${\bf u^0}$, is not 
trivial. Notice that because of the vectorial character of the field, the 
simple expression obtained in Ref. \cite{Armero98}
for scalar fields is
in general not applicable. Nevertheless, the corresponding eigenvector
can always be obtained at least numerically. 

Now, taking Eq. (\ref{deltafi}) and performing the scalar product with 
${\bf u^0}$, we obtain 
\be
\kappa ({\bf u^0},\hat{D}{\bf u_0}) + \beta \kappa ({\bf u^0},
{\bf u_0}) + \varepsilon^{1/2} ({\bf u^0},\boldmath{\mbox{$\Omega$}}
(\boldmath{\mbox{$\phi_0$}}; r,s;t) ) + \delta v(s,t) ({\bf u^0}, {\bf u_0}) = 0
\ee
where the scalar product is defined by
\be
({\bf f},{\bf g}) = \sum_i \int dr f_i(r) g_i(r)
\ee
Owing to the independence of the two first-order perturbations 
(curvature and fluctuations), we get 
\be
\beta(\varepsilon_\lambda) = - \frac{({\bf u^0}, \hat{D}{\bf u_0})} 
{({\bf u^0},{\bf u_0})} \label{beta}
\ee
and
\be
\delta v(s,t)= -
\varepsilon^{1/2} 
\frac{({\bf u^0},\boldmath{\mbox{$\Omega$}}(\boldmath{\mbox{$\phi_0$}}; r,s;t) )}
{({\bf u^{0}},{\bf u_0})}.
\label{deltav}
\ee
The stochastic process (\ref{deltav})
is not white since the 
high-frequency components of $\boldmath{\mbox{$\Omega$}}$ have been integrated 
out by the coarse-graining
procedure. However, we can now restore them harmlessly by replacing 
$\boldmath{\mbox{$\Omega$}}$ with the original multiplicative white 
noise process. By doing so, we are modifying the part which is 
not intended to be accounted for by the very eikonal description, while 
the treatment is simpler. Analogously, once the dependence on the cut-off
$\lambda$ has been explicitely worked out, 
we can take the process (\ref{deltav})
as delta-correlated in space. We will explicitly check the limit of validity 
of the 
eikonal description here proposed when scales comparable to the front thickness
are reached, in the sections below.

The resulting stochastic eikonal equation with the explicit dependence 
on the original noise parameters then takes the form
\be
v_n (s,t)= \bar{v}(\varepsilon_{\lambda}) +
\beta(\varepsilon_{\lambda}) \kappa(s,t) + 
D_f^{1/2}(\varepsilon,\varepsilon_{\lambda})\zeta(s,t) \label{jy}
\ee
where $\bar{v}(\varepsilon_{\lambda})$ is defined by Eq.(\ref{fi}),
$\beta(\varepsilon_{\lambda})$ is given by Eq.(\ref{beta}),
and the noise $\zeta(s,t)$ is a zero mean Gaussian white process with
\be
\langle \zeta(s,t) \zeta(s^{\prime},t^{\prime}) \rangle =
2 \delta (s-s^{\prime}) \delta(t-t^{\prime}),
\ee
which follows from the 
statistical properties of $\Omega$ with 
\be
D_f(\varepsilon,\varepsilon_{\lambda}) = 
\varepsilon \frac{\int dr \sum_{i,j} u_i^{0} u_{j(0)}
g_i(\boldmath{\mbox{$\phi_0$}})
g_j(\boldmath{\mbox{$\phi_0$}})}{({\bf u^0},{\bf u_0})^2}.
\label{tere}
\ee
Note that the dependence of $D_f$ on the dressed noise intensity
$\varepsilon_\lambda$ comes from the dependence on the same quantity of
$\boldmath{\mbox{$\phi_0$}}$ (and hence of ${\bf u_0}$ and ${\bf
u^0}$) as solution of the renormalized problem of Eq. (\ref{fi}).

The above equations constitute the main result of the first part of the
paper. 
Although the derivation is not rigorous because the coarse-graining could not 
be carried out explicitly, the result is appealing
from the theoretical point of view in that it separates the problem into 
an effective deterministic one, where the original field equations are 
modified by additional deterministic terms which depend on noise 
parameters, plus an additive noise which would be the necessary one to 
describe the wandering of the problem in one dimension. For the 
renormalized one-dimensional deterministic problem, therefore, the two present
perturbations, curvature and noise, decouple from each other. 

An important point to emphasize here is the separate dependence of the result on
two noise parameters, namely 
$\varepsilon$ and $\varepsilon_{\lambda}$. While the renormalized velocity
and the coefficient of the curvature depend solely on $\varepsilon_{\lambda}$,
the effective noise intensity $D_f$ depends separately on both. This 
clearly illustrates how the ultraviolet cutoff is an additional parameter of
the problem when the noise is multiplicative, in correspondence to the fact that
the continuum limit is not well defined for noise delta-correlated in
space. 
It is also important to remark that our derivation 
procedure is expected to be valid for small noise intensity $\varepsilon$, 
but contains all orders in $\varepsilon_{\lambda}$. This was already the case 
in the one-dimensional derivation of Ref.\cite{Armero98}, where the small 
noise approximation was phrased in terms of a separation of time 
scales. The connection between that scale separation, the coarse-graining 
procedure and the small noise expansion has been recently clarified in 
Ref.\cite{Rocco01}, where a rigorous derivation of the result of 
Ref.\cite{Armero98} has been presented for the case of a single-component field 
in one dimension, in terms of suitable projection 
techniques. 
Unfortunately, that rigorous derivation is based on the identification
of a specific collective coordinate which has no simple extension to higher 
dimensions. Nevertheless, the fact that this approximate procedure has
proven 
correct in $1d$ gives further support to our main result above, which is 
not claimed to be rigorously proven. 
In the following sections we will check this prediction
against analytical results and 
numerical simulations of the full reaction-diffusion equations in 
explicit examples. We will see that, the dependence on the cutoff $\lambda$ 
is essential not only for a quantitative description of the problem, but 
is crucial to predict nontrivial phenomena such the transition to pulled 
fronts, in which the whole eikonal description fails. 
This failure of the 
present description is signaled by the vanishing of the effective noise 
intensity $D_f$ when that point is reached. In fact, $D_f$ is linear in 
$\varepsilon$ to lowest order, but has a complicated dependence on 
$\varepsilon_{\lambda}$. 
As we will explicitly see, the partial resummation 
of orders in $\varepsilon_{\lambda}$ captures important physical features of the 
problem. For instance, it allows the 
non-monotonic dependence and eventual vanishing of the front diffusion coefficient
$D_f$. 
We expect the pushed-pulled transition to occur exactly at this point.
Another qualitative change captured by the above resummation is the destruction 
of the front itself, associated to the fact that the front thickness may diverge 
in some circumstances. This phenomenon is the signaled by a dicergence of $D_f$ 
at some finite value of $\varepsilon_{\lambda}$.

Although the predictions above are expected to be correct for 
$\epsilon_{\lambda} \sim 1$ as long as $\epsilon \ll 1$, in practice this 
may be limited by the fact that $\bar{v}(\varepsilon_{\lambda})$ and the 
Goldstone modes are not in general analytically known. 
In such case one can rely on numerical resolutions of the eigenvalue problem posed
by Eq. (\ref{fi}), or alternatively one can find $\bar{v}(\varepsilon_{\lambda})$ 
as a systematic expansion in powers of $\varepsilon_{\lambda}$ as described
in the Appendix. On the other hand, it is worth remarking that 
${\beta}$ can be a nontrivial function of $\varepsilon_{\lambda}$ only
for multicomponent fields, that is for pulses. For one-component fields 
(fronts) the coefficient $\beta$ is not renormalized by noise. 

We remark that this derivation is
meaningful only when pulses are involved or, in the case of fronts,
when the relevant dynamical regime is the
pushed one. In the last Section of this paper, we shall give more
details about the main differences between pushed and pulled fronts,
and about their different response to noise. For the time being, we
point out that the lack of time-scale separation between the
relaxation of the zero mode and the other eigenmodes of the 
spectral operator for pulled fronts prevents in general from
constructing a local equation for the interface motion \cite{Ebert00b}. 
Examples of reaction-diffusion systems which do not admit local kinematic 
descriptions are the phase-field formulations of solidification 
\cite{Wang93,cinca} or viscous fingering \cite{folch}. In those cases a nonlocal
interface equation does exist so an extension of our derivation 
is in principle feasible. This may be particularly interesting in cases such
as in Refs. \cite{cinca,folch} were the relevant fluctuations may be external.

\section{Kinetic roughening and connection to equilibrium fluctuations}

\begin{figure} 
\centerline{{\psfig{figure=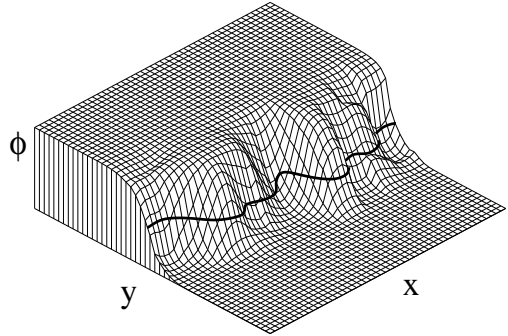,angle=-90,height=2.4in}}}
\caption{\small{An example of a noisy front with a level curve which 
defines the precise location of the front.}} \label{front}
\end{figure}

In the previous Section, we have derived 
a stochastic sharp interface
approximation for a generic RD system with multiplicative noise. 
A pictorial description of a noisy front  
is given in Fig. \ref{front} 
We see in this figure that noise in the RD system induces 
fluctuations in the front
shape, thus generating roughening of the sharp interface 
that should emerge at the eikonal
level of description. The identification of universality 
classes of kinetic roughening will come naturally at this level of description.
In particular we will establish the connection with the universality classes 
defined by the Kardar-Parisi-Zhang (KPZ) Equation \cite{Kardar86}
and by the Edwards-Wilkinson (EW) Equation \cite{Edwards82}.

The stochastic eikonal equation (\ref{jy}) is written in instrinsic, rotation 
invariant form. For the purposes of scaling theory it is convenient  
to write it in cartesian coordinates. The front location is then given as 
$y=h(x,t)$. Retaining only the relevant nonlinear terms in the Renormalization 
Group (RG) sense, we then recover the KPZ equation
\be
\frac{\partial h}{\partial t} = \nu \frac{\partial^2 h}{\partial x^2}
+ \frac{\lambda}{2} \left(\frac{\partial h}{\partial x}\right)^2 + \mu(x,t),
\ee
with
\be
\langle \mu(x,t) \mu(x^{\prime},t^{\prime}) \rangle = 
2 D_{\rm KPZ} \delta (x-x^{\prime}) \delta(t-t^{\prime}). \label{corrkpz}
\ee
The KPZ parameters turn out to be related to the eikonal ones as
\be
D_{\rm KPZ} = D_f, \;\;\;\;\;\;\;\; \nu = \beta, \;\;\;\;\;\;\;\; 
\lambda = \bar{v}. \label{id}
\ee

In the special case of $\bar{v}=0$, the EW equation is obtained,
\be
\frac{\partial h}{\partial t} = \nu \frac{\partial^2 h}{\partial x^2}
+ \mu(x,t), \label{EW}
\ee
with
\be
\langle \mu(x,t) \mu(x^{\prime},t^{\prime}) \rangle = 
2 D_{\rm EW} \delta (x-x^{\prime}) \delta(t-t^{\prime}), 
\ee
and again 
\be
D_{\rm EW} = D_f, \;\;\;\;\;\;\;\; \nu = \beta. \label{ident}
\ee
These equations are well known to be the paradigm for many different
growth processes \cite{stanley,Krug97}. 
Even if the microscopic dynamics of the system under
study may correspond to different equations of motion for the
respective interfaces or surfaces,
nevertheless the KPZ and EW Equations do capture the universal 
features of the system, namely the scaling properties of fluctuations.

Usually, such effective equations cannot be derived from the original 
microscopic description of the particular systems and are introduced on 
a phenomenological basis, relying on the claim of universality within 
a RG framework. Nonuniversal quantities such as prefactors of scaling 
functions, affected for instance by the noise intensity in the interface 
equation cannot be derived. In our case, we are able to compute the noise 
intensity in the eikonal equation so we can also predict the nonuniversal 
prefactors if the noise is known in the reaction-diffusion level of description.
%%@@
For instance, not only the scaling of the interface roughness with 
system size can be predicted, but also the actual values of average interface 
roughness in terms of the original microscopic parameters of the RD model
are worked out.

As a test of our derivation we will now check consistency with equilibrium 
fluctuation theory. The connection between bulk thermal fluctuations 
and fluctuations of the interface between thermodynamical phases can be
established rigorously in the case of equilibrium fluctuations. 
This is possible because a free energy functional 
does exist and the sharp-interface limit can be performed at the level of the
free energy itself. 
Then, the fluctuations can be obtained independently from the free energy 
at each level of description (either bulk or interface fluctuations), 
consistently with the fluctuation-dissipation theorem.
The important difference here is that no dynamical
equation must be invoked but only equilibrium properties. 
On the contrary, in the more general case
where
there is no fluctuation-dissipation theorem and not even a free energy 
functional, we must rely on dynamical equations. In the case of equilibrium
fluctuations, however, we must reproduce the 
known correct result.
As we will see below, this case falls in the EW universality class. 

The general solution of the EW Equation, Eq. (\ref{EW}), is known \cite{Krug97}.
Consider the interface $h(x,t)$ and its discrete Fourier Transform 
$\hat{h}_q(t)$, defined through
\be
h(x,t) = \sum_q  \hat{h}_q(t) \exp(iqx).
\ee
It is possible to show \cite{Krug97} that the long time limit of the spectrum
$S(q,t) = \langle |\hat{h}_q(t)|^2 \rangle$ is given by the expression
\be
\lim_{t \rightarrow \infty} S(q,t) = \frac{1}{L} \frac{D_{\rm EW}}{\nu q^2}.
\label{lims}
\ee
Our strategy now will be to calculate explicitly the spectrum 
(\ref{lims}) in terms of the coefficients predicted for 
our eikonal equation, and
then show that the resulting expression coincides with the independent
result that can be obtained from equilibrium fluctuation theory.
Hence we insert an additive noise
in the original RD system, that is in setting 
$g=1$:  
\be
\frac{\partial \phi}{\partial t} = D \nabla^2 \phi + 
F(\phi) + \varepsilon^{1/2} \eta({\bf x},t),
\label{rdv0}
\ee 
with
\be
\langle \eta({\bf x},t) \eta({\bf x}^{\prime},t^{\prime}) \rangle = 
2 \delta({\bf x}-{\bf x}^{\prime}) \delta(t-t^{\prime}). 
\ee
We consider a $F(\phi)$ with a symmetric double-well form, 
{\it i.e.}
the deterministic part of Eq. (\ref{rdv0}) has a $1d$ solution 
with zero velocity
$\phi=\phi_0(x)$. This is what corresponds to the usual time-dependent 
Gizburg-Landau Langevin equation for a nonconserved order parameter 
(Model A in the Hohenberg-Halperin classification \cite{Hohenberg}), where 
noise intensity must be identified as $\varepsilon=k_BT$. 
Since $g=1$, our expression for the noise intensity at the sharp interface 
level takes the simpler form 
\be
D_f=\frac{\varepsilon}{\displaystyle \int_{-\infty}^{\infty}dx\left
(\frac{\partial \phi_0} {\partial x}\right)^2},
\ee 
which, according to (\ref{lims}) and performing the identifications of Eq.
(\ref{ident}) with $\beta=D$, produces  
\be
\lim_{t \rightarrow \infty}
\langle |\hat{h}_q(t)|^2 \rangle = \frac{\varepsilon}{L D \int_{-\infty}
^{\infty}dx\left (\frac{\partial \phi_0} {\partial x}\right)^2} \frac{1}{q^2}.
\label{resu}
\ee
On the other hand, we can take the sharp-interface limit on the free 
Ginzburg-Landau free energy. The calculation is standard 
(see for instance Ref.\cite{Safran94}) and yields the interface free  
energy
\be
F_I=\sigma \int dx \sqrt{1+\left(\frac{\partial h}{\partial x} \right)^2},
\ee
where the parameter $\sigma$ is identified as the interfacial tension
and can be evaluated from the bulk free energy of the system \cite{Safran94}
as
\be
\sigma = D \int dx \left(\frac{\partial \phi_0}{\partial x} \right)^2,
\label{sigma} 
\ee
where $\phi_0$ is the corresponding kink solution. 
For soft (long wavelength) deformations of the interface, 
the excess free energy reads
\be
\Delta F_I \approx \frac{\sigma}{2} \int dx \left(\frac{\partial h}
{\partial x} \right)^2,
\ee
and the corresponding stationary spectrum of fluctuations, consistent with 
the fluctuation-dissipation theorem takes the form 
\cite{Safran94}  
\be
\langle |\hat{h}_q|^2 \rangle = \frac{k_B T}{L \sigma q^2}.
\ee
Using Eq.(\ref{sigma}), this expression yields the same 
result of Eq.(\ref{resu}). 
This proves that the front roughening obtained from our derivation 
is exact in the case of equilibrium fluctuations, and by extension
in the additive noise case.
We have shown this
for the case of a non-conserved order parameter. In the conserved case (Model 
B of Ref.\cite{Hohenberg}), the projection to a sharp interface
description yields 
a non-local equation and therefore lies outside the validity of our theory. 
Similarly, equilibrium fluctuations have also been studied in both 
sharp-interface \cite{karmaprl} and phase-field \cite{rappel,cinca} 
formulation in the context of solidification,
which also yields non-local interface dynamics. The universality classes in 
those cases are not well established.

\section{Application to a prototype model of front propagation}

To illustrate our general theory we 
consider here as an example the propagation of a scalar front 
\be
\frac{\partial \phi}{\partial t} = D \nabla^2 \phi +
F(\phi,a) + \varepsilon^{1/2} g(\phi) \eta, \label{eqmain}
\ee
with the noise correlator defined as in Eq. (\ref{noisecorr}).
We specify our prototype model through the following definitions:
\be
&F(\phi)=\phi(1-\phi)(\phi+a) \label{proto} \\
&g(\phi)=\phi(1-\phi), \label{protog}
\ee
and we will consider a front of the $\phi=1$ state invading the $\phi=0$ one. 
The constant $a$ is a control parameter. 
As it is well known for fronts without fluctuations, 
the deterministic force given
by (\ref{proto}) leads to different modes of front propagation depending on the 
value of $a$ \cite{vanSaarloos}. The front velocity depends also on $a$.
The choice of the coupling function (\ref{protog}) for the multiplicative noise
term of Eq. (\ref{eqmain}) is the simplest one that preserves the 
stationary states $\phi=0$ and
$\phi=1$. The fact that the noise term vanishes in the two asymptotic states
prevents nucleation phenomena in the invaded state. The form of $g(\phi)$ is 
also such that the multiplicative noise term arises naturally as the 
fluctuation of the control parameter $a$. Moreover, for the prototype model 
here proposed, the corresponding function $h(\phi)$ defined in 
Eq.(\ref{functh}) which appears in the renormalized equation takes the same 
functional form as $f(\phi)$, only with renormalized coefficients. This 
allows a simpler analytical treatment and intepretation of the 
results.

\subsection{The 1d case revisited}

This prototype model has already been analyzed 
for 1d in Refs.
\cite{Armero96,Armero98}, 
where in the regime $ -\varepsilon_\lambda < a < 1/2-\varepsilon_\lambda$ 
it is proven to result in a renormalized average velocity 
\be
\bar{v}=\frac{2a + 1}{\sqrt{2(1-2\varepsilon_\lambda)}} \label{vavera}
\ee
and in the diffusion coefficient
\be
D_f = \varepsilon \;\frac{\int_{-\infty}^{\infty} d \xi \;
e^{2 \bar{v} \xi}\;(d \phi_0/d\xi)^2 \;g^2(\phi_0)}
{\left[\int_{-\infty}^{\infty} d \xi \;e^{\bar{v} \xi}
\;(d\phi_0/d\xi)^2\right]^2}. \label{formula}
\ee
Here $\phi_0$ is the
solution of the zeroth order equation:
\be
\frac{d^2 \phi_0}{d \xi^2} + \bar{v} \frac{d \phi_0}{d \xi} +
h(\phi_0)=0. \label{partialeqm}
\ee
Notice the dependence of both these quantities on $\lambda$ through
the $\varepsilon_\lambda$ parameter. 
Specifically about the diffusion coefficient, the
first order dependence in $\varepsilon$ is apparent, while the
functions present in the integral and defined through
(\ref{partialeqm}) contain all orders in $\varepsilon_\lambda$. This is in 
contrast to the renormalized velocity, which depends solely on 
$\varepsilon_\lambda$. This means that, both $\varepsilon$ and $\lambda$ 
can be determined independently from separate measurements 
of both the ballistic and the
diffusive components of the front propagation. This is quite remarkable since 
it provides indirect means of measurement of the (microscopic) noise, 
which may not be directly accessible in many cases.

\begin{figure} 
\centerline{{\psfig{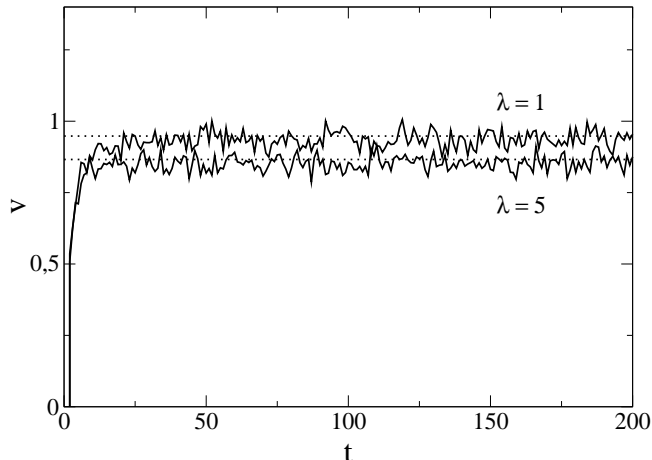}}}
\caption{\small{Change with $\lambda$ of the average renormalized
velocity. Values of the parameters are: $a = 0.1$, $\varepsilon=0.1$,
$\lambda$=1,5. The theoretical values of the corresponding average
velocities as from Eq. (\ref{vavera}), 0.948 and 0.866, are also
plotted.}} \label{figlambdavel}
\end{figure}

As an illustrative and well-controlled example, we have explicitly tested 
the prediction of the dependence
on $\lambda$ of both velocity and diffusion coefficient with direct numerical 
simulation of the RD equation.  
The first set of results is shown in Fig. \ref{figlambdavel}. Here
data from a simulation of the RD equation with the reaction 
term (\ref{proto}) are reported and compared with the theoretical
prediction (\ref{vavera}) for two different values of $\lambda=1$  
and $\lambda=5$, in some dimensionless units of the simulation. 

Notice that the result is indeed sensitive to the 
microscopic cutoff $\lambda$ of the noise, even though this length is 
significantly smaller
than the front thickness, of the order of 25, in the same units.

\begin{figure} 
\centerline{{\psfig{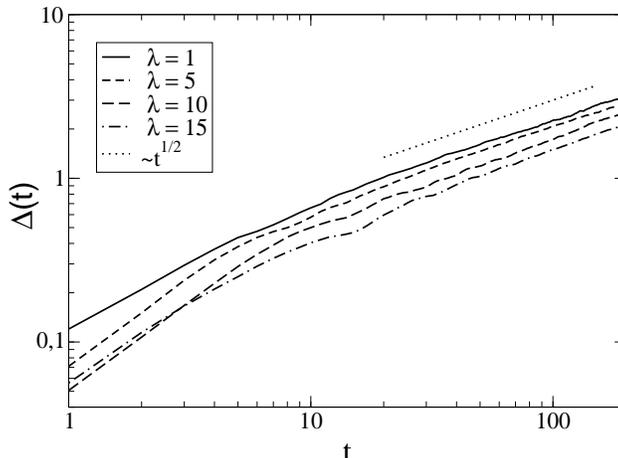}}}
\caption{\small{Change with $\lambda$ of the diffusion
coefficient for the front wandering in 1d. Parameters have the same 
values as in Fig. \ref{figlambdavel}.}}\label{figlambdadiff}
\end{figure}

Regarding the diffusion coefficient (\ref{formula}), we measured the mean
square displacement of the front position. If we define the front
position as 
\be
z(t) = \int_{x_0}^{\infty} dx \phi(x,t),
\ee
then the diffusion coefficient (\ref{formula}) is related to the mean
square displacement 
\be
\Delta = \sqrt{\langle z^2 \rangle - \langle z \rangle^2}
\ee
as 
\be
\Delta^2(t) \sim 2 D_f t
\ee
In Fig. \ref{figlambdadiff}  the quantity $\Delta(t)$ is plotted.
The values of the parameters are the same as in Fig. \ref{figlambdavel},
while the values of $\lambda$ run from 1 to 15. The diffusive
behavior and the dependence on the value of $\lambda$ is manifest. 

In this particular case of 1d, a more systematic derivation of the diffusion 
coefficient has been reported in Ref.\cite{Rocco00}. 
By proper identification of the
natural collective variable which describes the front wandering as strictly 
(not just asymptotically) diffusive, it has been 
shown that, in fact, the result first found in Ref.\cite{Armero98} is rigorous
to first order in $\varepsilon$. Together with the case of equilibrium 
fluctuations, this is the second rigorous test of our general theory.

\subsection{The 2d case}

Let us now consider the propagation of a front in the prototype model of Eqs.
(\ref{eqmain}, \ref{proto}, \ref{protog}) in 2d. 
We have already shown that in two dimensions the eikonal equation
reduces to either the EW equation or the KPZ equation, depending on 
whether the velocity of a planar front is zero or non-zero respectively. 
In order to make a numerical check of our theory we consider then 
the simplest case of zero velocity, with expected EW scaling.
This case corresponds to the choice $a=-1/2$ in Eq. (\ref{proto}). 
As seen explicitly in Eq.(\ref{vavera}) it turns out that the 
renormalized velocity is
also zero, since the noise does not break the symmetry associated to the 
double-well form of the deterministic potential. It is important to remark 
that, unlike the Ginzburg-Landau model discussed in Section III for the 
equilibrium case, the noise is now multiplicative, and no 
fluctuation-dissipation theorem can be invoked. 
Hence the first-principles derivation of the fluctuation spectrum in the 
sharp-interface description is no longer available.

We thus rely on the dynamical equation   
\be
\label{sym}
\frac{\partial \phi}{\partial t} = D  \nabla^2 \phi + \phi(1-\phi)
(\phi-\frac{1}{2}) + \varepsilon^{1/2}\phi(1-\phi) \eta({\bf x},t), 
\ee
where the noise $\eta$ is defined through the usual correlator of 
Eq.(\ref{noisecorr}). 

Now, our basic goal is to connect this level of description with the
eikonal level, determining thereby the noise corrections
to this equation. Therefore we  
assume a stochastic eikonal equation of the form
\be
\left\{
\begin{array}{l}
v_n(s) = - \beta \kappa + D_f^{1/2} \zeta(s,t)\\ 
\langle \zeta(s,t) \zeta(s^{\prime},t^{\prime}) \rangle = 
2 \delta (s-s^{\prime}) \delta(t-t^{\prime})
\end{array}
\right.
\ee
where the coefficients $\beta$ and $D_f$ are given by
\be
\beta = D, \;\;\;\;\;\;\;\;\;\;
D_f = \frac{\int dx u^{0} u_{0} g^2(\phi_0)}{(u^{0},u_{0})^2}.
\label{betaD}
\ee
Notice that we have directly taken into account that the bare as well as
the noise-renormalized velocities are both zero 
(see for example \cite{Armero98}). Also notice that the
renormalization of the curvature term is absent for 
both $\varepsilon$ and $\varepsilon_\lambda$ expansions due to
the fact that we are now dealing with a front ($N=1$).

Now, in order to calculate $D_f$, we need to specify the solution of
the 1d model. This is known and for $a=-1/2$ is given by
\be
\phi_0 = \frac{1}{2}(1-\tanh kx), \;\;\;\;\;\;\;\;\;\;\; k=\frac{1}{2 \sqrt{2}},
\ee
and
\be
u^{0} = u_{0} = \frac{d \phi_0}{dx}.
\ee 
The integral in (\ref{betaD}) can then be computed exactly and gives:
\be 
\label{exact}
D_f = \varepsilon \frac{9}{35} \sqrt{\frac{2}{1-2 \varepsilon/\lambda^2}}. 
\ee 
The above result clearly illustrates the different treatment of the parameters 
$\varepsilon$ and 
$\varepsilon/\lambda^2$. The result is first order in $\varepsilon$ and 
contains 
all orders in $\varepsilon/\lambda^2$. It is interesting to remark that the
partial resummation of all orders in $\epsilon$ captures important 
nonpertubative phenomena. For instance, the divergence of $D_f$ at 
$\varepsilon/\lambda^2=1/2$, reflects the fact that at this point the 
front itself is destroyed, or, equivalently, the front thickness becomes 
infinite. This is equivalent to reaching a critical point, except that this 
is not the equilibrium one because the noise is multiplicative. 
Remarkably, 
our result for additive noise case does not capture that feature
because it is only first order in the noise strength.
In the case with an asymmetric double-well potential, which has a finite front
velocity, it was explicitly checked numerically in Ref.\cite{Armero98} that
the diffusion coefficient of the front has a non-monotonic dependence with
$\varepsilon/\lambda^2$. Most importantly it vanishes at a finite value of 
$\varepsilon/\lambda^2$ which corresponds exactly to the point were the front 
reaches the pushed-pulled transition. Again we see that the resummation of 
orders $\varepsilon/\lambda^2$ captures important physical information 
(see discussion in Section V).

We now come back to the numerical test of the EW scaling of our particular 
symmetric case, with the identification (\ref{ident}). 
Accordingly, we can rewrite the complete power spectrum as it is known
theoretically \cite{Krug97} in terms of the coefficients that we just
calculated, and compare it with data from a direct simulation of the field 
model Eq.(\ref{sym}).

From \cite{Krug97} the power spectrum $S$ reads, 
\be
S(q,t) = \frac{1}{L} \frac{D_{\rm EW}}{\nu q^2}(1-\exp(-2 \nu q^2 t)). 
\label{powspectr}
\ee
In terms of the parameters in the original field equation this yields 
\be
S(q,t) = \frac{1}{L} \frac{D_{f}}{D q^2}(1-\exp(-2 D q^2 t)),
\label{predicted} 
\ee
where $D_f$ is given by Eq.(\ref{exact}). Accordingly, the fluctuations of 
the front position in the RD model Eq.(\ref{eqmain}) for length scales 
larger than the front thickness itself must obbey the spectrum defined 
by Eq.(\ref{predicted}). It is worth remarking that the prediction is not 
only for the universal features, namely the shape of the scaling function 
and the exponents, but for the actual absolute values of the spectrum. 
We have performed simulations of the field RD model with a correlator of
the form
\be
\langle \eta({\bf x},t) \eta({\bf x}^{\prime},t^{\prime}) \rangle =
\frac{1}{\lambda^2}
(1-|{\bf x}-{\bf x}^{\prime}|/\lambda)
(1-|\vec{y}-\vec{y}^{\prime}|/\lambda)
\Theta (1-|{\bf x}-{\bf x}^{\prime}|/\lambda)
\Theta (1-|\vec{y}-\vec{y}^{\prime}|/\lambda)
\delta(t-t^{\prime})
\ee
This corresponds to assuming that at any time, the noise takes the same 
value in a square of side $\lambda$, uncorrelated with the neighboring 
squares. This is done for simplicity but no significant dependence is expected 
on the details of the shape of the spatial correlation provided $\lambda$ is
kept smaller than front thickness.

\begin{figure} 
\centerline{{\psfig{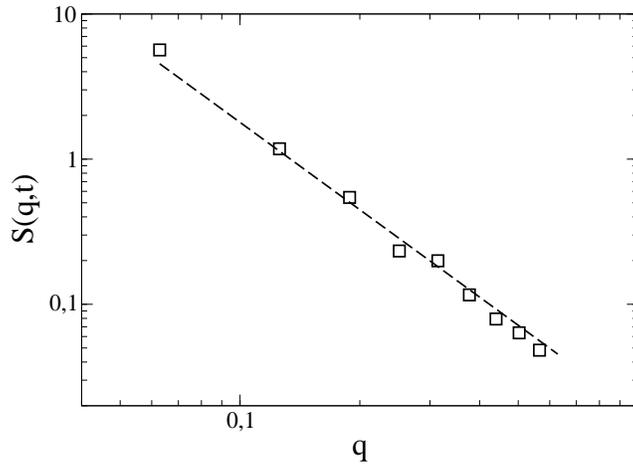}}}
\caption{\small{Numerical data from 2d simulation of the RD 
field equations for the protorype model 
with $a=-1/2$ and analytical prediction for the
power spectrum (\ref{powspectr}) in the scaling region.
The time is here 1000, and the parameters of the simulation are
$L=100$, $\varepsilon=5$, $\lambda=5$.}} \label{figspectr1}
\end{figure}

\begin{figure} 
\centerline{{\psfig{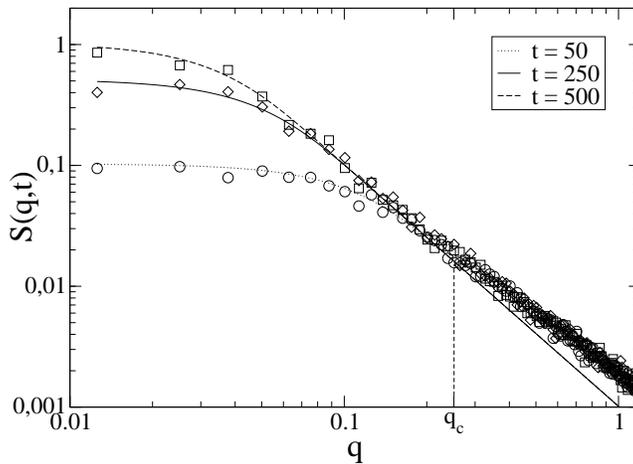}}}
\caption{\small{Numerical data from 2d simulation of the RD field 
equations for the protorype model with $a=-1/2$ and analytical prediction 
for the power spectrum (\ref{powspectr}). The three sets of data points refer
to times 50, 250 and 500. The parameters of the simulation are
$L=500$, $\varepsilon=1$, $\lambda=2$. The value of $q_c$ corresponds to 
a wavelength of the order of the front thickness. The analytical prediction 
is only intended for $q < q_c$.}} \label{figspectr2}
\end{figure}

We have studied the fluctuations of the internal level curve of the front 
$\phi=1/2$.
In Fig. \ref{figspectr1} we show the scaling region, with the correct slope 
and location of the curve. More remarkably, Fig. \ref{figspectr2} shows
the measured spectra for the simulation of the field equations compared to 
the prediction given by Eq.(\ref{predicted}). 

It should be stressed that in this comparison there is no free parameter.
It is also interesting to observe the deviations from the prediction at 
length scales smaller than the front thickness. In this high-q region the 
data also collapse but not to EW scaling. An estimate of the exponent 
$\alpha$ in this region is around $3/2$.

For values of $a$ such that the front has a finite velocity, one would 
expect that the scaling would be given by that of the KPZ universality 
class. Although the scaling function in that case is not exactly known, 
the prediction of our eikonal equation is expected to fit the data for
the corresponding RD model also without free parameters. We have not checked 
this case because it is obviously more involved and less conclusive 
because of the practical difficulties to reach the scaling regime already 
at the eikonal level of description \cite{Beccaria94}. 

From a practical point of view it is to be remarked that the noise intensity 
cannot be increased arbitrarily in a simulation without destroying 
the front itself. 
This can be easily seen in a numerical simulation. 
Although the noise vanishes asymptotically in the 
stationary states $\phi = 0$ and $\phi = 1$, if noise is sufficiently strong
it may be capable to nucleate the other state in the region not too far from 
the front. We have found that this effect is more pronounced in the 
region behind the front $\phi = 1$. For a given $\lambda$ there will typically
be a maximum value of $\varepsilon$ after which the front is essentially 
destroyed. If we increase $\lambda$ the effect is milder. In order to 
make the front roughening appreciable in not too large system sizes, it is 
thus convenient to have a moderately large $\lambda$, which in turn will 
allow larger values of $\varepsilon$. Typical values that we have 
considered are in the range of $\lambda=2,5$ n units for which the front thickness is of the
order of $25$.

\section{The pushed - pulled transition}
As we have being mentioning so far, two classes of fronts must be 
distinguished from a dynamical point of view, the so-called 'pushed' and
'pulled' fronts \cite{Ebert00a,Ebert00b}.
The simplest is pushed case, in which 
the front propagation  
depends on the full non-linear structure of the equation of
motion and the front is said to be 'pushed' by its internal part. This is 
usually the case when the invaded state is locally stable. 
On the contrary, if the invaded state is unstable, it can happen that 
the relevant dynamics takes place in the semi-infinite 
leading edge region ahead of the front itself. Then the
propagation of the front is governed by the growth and spreading of
linear perturbations in that region which 'pulls' the front. In this case 
the linearization about the
unstable state accounts for its dynamical behavior \cite{Ebert00a}, but 
there is degeneracy of propagating velocities\cite{vanSaarloos}. In the 
present context, the most important distinction between the two situations 
is that, while for pushed fronts, the relaxation of bulk modes is exponential, 
for pulled fronts it is algebraic, as a result from the fact that the 
linearized operator describing perturbations around the stationary propagating 
mode is gapless. This means that for pulled fronts, there is no natural 
time-scale separation which allows for the decoupling of the interface modes 
from the bulk ones. Our derivation, and the whole idea of a kinematic 
moving-boundary approximation, is not valid for pulled fronts\cite{Ebert00b}.
Nevertheless, we will show that our theory does predict correctly its 
failure at the pushed-pulled transition. 

The intrinsic differences between pushed and pulled fronts also show 
dramatically in the statistics of fluctuations. In 1d, for instance the 
front wandering turns from difussive (pushed fronts) to subdiffusive 
(pulled fronts) \cite{Armero98,Rocco00}. In turn, pulled fronts in 2d with 
multiplicative noise have been found to belong to a different universality 
class than the ordinary KPZ \cite{Tripathy00,Tripathy01}, as opposed to
the KPZ scaling of pushed fronts.

\begin{figure} 
\centerline{{\psfig{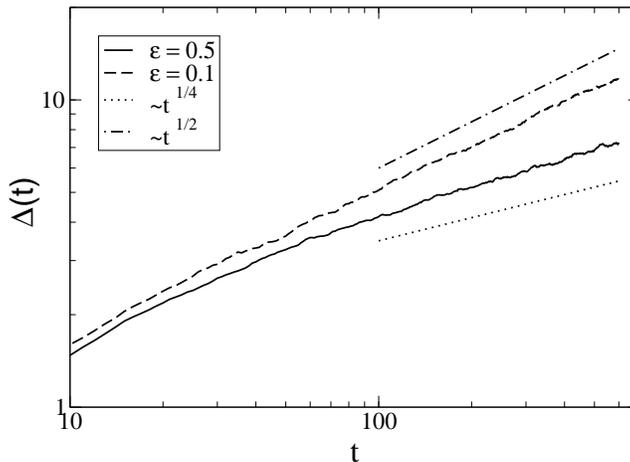}}}
\caption{
\small{The pushed-pulled transition in 1d. The system size is $L=2000$ and
averages have been carried out on $3600$ realizations of noise in
the pulled case and about $1000$ in the pushed one.}}
\label{transition}
\end{figure}

In Ref.\cite{Armero98} it was already observed that the diffusion coefficient 
of fronts in 1d was a nonmonotonic function of $\varepsilon/\lambda$ which
vanished at a finite value of that parameter. After that point, subdiffusive
behavior was found. The picture was confirmed and completed in Ref.
\cite{Rocco00}. Here we want to stress that the point where that transition 
occurs corresponds {\it exactly} to the pushed-pulled transition. In fact,
the renormalized equation defined by Eq.(\ref{main}) for the protoype model 
takes the same form as the original one, with renormalized coefficients. 
The deterministic equation has the transition to the pulled regime at 
$a=1/2$. The same transition in the renormalized equation does occur at 
$\tilde{a}=1/2$ where $\tilde{a}= a + \varepsilon_\lambda$. It is thus clear 
that, for parameter values of the deterministic equation in the pushed regime,
increasing noise intensity $\varepsilon$ or decreasing $\lambda$ will 
imply a transition to the pulled regime. While the front velocity will not 
experience any dramatic effect, the fluctuations (wandering in 1d or 
roughening in 2d) will be dramatically affected, since the universality 
class will change. In Fig. \ref{transition} we show the change from 
diffusive to subdiffusive behavior induced solely by a change in the 
effective noise instensity $\varepsilon_\lambda$.
Once more it is remarkable the dynamical importance of the 
noise correlation length $\lambda$ in the long-wavelength behavior of 
the front. The same effect should be expected in 2d, namely, the scaling 
could be KPZ or non-KPZ depending solely on noise parameters. 
To our knowledge, this is the first time that such a dramatic effect of
noise is reported.
Remarkably enough, while our eikonal description is not able to describe 
the second regime, it does predict the transition at the right values of
parameters.

\section{Discussion and Conclusions}

The formulation  
in terms of kinematic eikonal-like equations provides a useful
framework for studying front or pulse propagation when one is interested in long 
spatial and temporal scales.
This kind of equations have advantages both for numerical
simulations and for theoretical analysis, and has fruitfully been used for instance 
in the iphenomenological modeling of excitable wave 
propagation in disordered and noisy media, by the \it ad hoc\rm  procedure of adding
fluctuations to a generic eikonal equation. In this paper we have derived
stochastic eikonal equations from the more microscopic RD 
field equations with noise, which
can be multiplicative. 

The derivation presented here relies on the hypothesis of separation of scales
between the front dynamics at large scales and the internal degrees of freedom of
the front. That means that it is not valid for pulled fronts, which are indeed
known not to have a local, eikonal-like description even in the absence of noise.
However, even for the pushed case, the usual projection techniques for derivation 
of sharp-interface equations 
cannot be simply extended to the stochastic case due to the fact that the noise 
contains the full range of length and time scales. Hence our derviation is formulated 
within a coarse-graining scheme, which we claim makes sense for the kind of problems 
which admit a local, eikonal-like equation. Derivation of sharp-interface approximations 
with noise have been possible so far for thermal noise only, 
in systems with local equilibrium.
For multiplicative, generically external noise, however, the absence of a free energy 
and a fluctuation-dissipation theorem requires an alternative scheme based on 
dynamical equations. We have explicitly checked that our scheme is exact for
the cases of equilibrium fluctuations and also reproduces a rigorous result for 
multiplicative noise in d.
For the general case, however,
we rely on numerical tests to fully justify 
its validity. In any case, the excellent agreement with numerical and 
analytical tests clearly suggests that a more rigorous derivation should be
possible. An extension of our procedure is also conceivable in RD systems for which 
a nonlocal sharp-interface description exists for instance in solidification or viscous
fingering. 

One of the main points of this paper 
has been to clarify the role of the spatial cutoff of noise 
correlations $\lambda$. On the basis of recent rigorous mathematical findings on 
the spatio-temporal white-noise limit, we have argued and shown in explicit examples,
 that the common splitting  
of the Stratonovich white noise in a term which is singular as $\lambda \rightarrow 0$ 
plus an It\^o noise, does capture the correct dependence 
on $\lambda$ of macroscopic 
quantities (at scales much larger than $\lambda$). The remaining spatially-correlated 
It\^o noise will only carry a weak dependence on $\lambda$, noticeable at length 
scales of the order of $\lambda$, and can be treated perturbatively. 
The $\lambda \rightarrow 0$ limit can thus be taken safely for the 
remaining It\^o noise, once 
the singular part has been extracted.
Within this scheme, 
macroscopic quantities such as front velocity or front roughening properties, 
depend separately of the noise strength $\varepsilon$ and on $\varepsilon/\lambda^d$. 
Incidentally, this implies the possibility of measuring the (microscopic) 
noise parameters $\varepsilon$ and $\lambda$ from the macroscopic
 dynamics of the fronts.
Most remarkably, a partial resummation scheme can be naturally defined for the 
parameter $\varepsilon/\lambda^d$
which captures some important, nonperturbative physical phenomena which would be missed 
otherwise, such as the transition from pushed to pulles regimes, or the destruction 
of the front associated to a divergence of the front thickness. These phenomena,
at the edge of validity of the sharp-interface approximation, are detected respectively 
by the vanishing and the divergence of the effective noise-strength on the eikonal 
equation.

Eikonal stochastic equations like the ones derived here are directly related to the
EW and KPZ equations of kinetic roughening. By using known results for these equations and
the results presented here one can predict roughening properties of noisy pulses
or fronts appearing in field equations. We have used such correspondence to check
the predictions in a case of zero velocity and additive noise, for
which the general
theory of equilibrium fluctuations can be directly applied to the RD equation.
Analytical results obtained for the corresponding (Edwards-Wilkinson) eikonal
equation have been identical (including prefactors) to that independent
calculation for the RD system.

We have also applied our results to a prototype model with multiplicative noise
constructed to represent a variety of different front propagation regimes by
changing a single control parameter. While this model has already been used to
study effects of noise on $1d$ fronts, we have addressed here a more specific $2d$
effect such as the case of front roughening. In a zero-velocity case, simulations
of the reaction-diffusion equation have presented a very good agreement with the
predictions of roughening for the corresponding EW equation with no adjustable parameter. 
The results show that the dependence on $\lambda$ is quantitatively important 
even if $\lambda$ is significantly smaller
than the front thickness, which may seem counterintuitive. 
Although we have not
checked the analogous results for a nonzero velocity, which would correspond to
the KPZ equation, we expect our results to be valid also in this case.

Most interestingly we have explicitly checked the prediction of qualitative changes 
as $\varepsilon/\lambda^d$ is varied, such as the transition from pushed to pulled 
propagation regimes. We have directly observed the change in the wandering exponent in 
1d as $\lambda$ is decreased. The immediate extension of this result implies that 
a transition from KPZ to non-KPZ scaling is to be expected in higher dimensions.
We thus conclude that the dependence on the spatial cutoff of 
the noise may have dramatic effects, not only on the nonuniversal quantities but also 
on the universal ones.

In summary, we have derived stochastic eikonal equations from stochastic RD
equations completely specifying the parameters of the noise. Although this is not
a systematic derivation, we have presented a wealth of independent 
evidence on the validity of our results both from analytical
calculations and from numerical simulations of different systems and looking at
distinct noise effects, both qualitative and quantitative. 
These results should be of interest for theoretical and
practical purposes both in the study of kinetic roughening and in the context of
propagation of chemical waves.

\acknowledgments 

We acknowledge financial support from Direcci\'on
General de  Investigaci\'on Cient\'{\i}fica y T\'ecnica (Spain)
(Projects BFM2000-0624-C03-02 and BXX2000-0638-C02-02),
Comissionat per a Universitats i Recerca (Spain)
(Projects 1999SGR00145 and 2000XT-0005),
and European Commision (Project TMR-ERBFMRXCT96-0085). We also
acknowledge computing support from Fundaci\'o Catalana per a la
Recerca-Centre de Supercomputaci\'o de Catalunya (Spain).

\section*{Appendix: Systematic Expansion}

In some circumstances, the eigenvalue problem defined by Eq. (\ref{fi}) may not be
solvable analytically, due to the functional form of
${\bf h}(\boldmath{\mbox{$\phi_0$}})$, which has been renormalized by noise, even though 
the 'bare' problem may be solvable.
In such cases it may be useful 
to solve for the case $\varepsilon = 0$ (no noise) and
obtain corrections in a systematic expansion in $\varepsilon$. This is the aim of
this appendix.

Let us consider then the zero noise eigenvalue problem defined by the
(deterministic) $1d$ equation
\be
0 = \hat{D} \frac{\partial^2 \boldmath{\mbox{$\phi_{\cal D}$}}}{\partial r^2} 
+ v_0 \frac{\partial \boldmath{\mbox{$\phi_{\cal D}$}}}{\partial r} +
{\bf f}(\boldmath{\mbox{$\phi_{\cal D}$}})  \label{fi0}
\ee
We now expand the fields of the noisy 2d problem of Eq. (\ref{wtho}) as
perturbations in $\varepsilon$ and $\kappa$ of the solution of Eq. (\ref{fi0}):
\be
\boldmath{\mbox{$\phi$}}(r,s,t) = \boldmath{\mbox{$\phi_{\cal D}$}}(r) 
+ \delta\boldmath{\mbox{$\phi$}}(r,s,t)
\label{pe0}
\ee
\be
v_n(s,t) = v_0 + \alpha \varepsilon
+ \beta \kappa(s,t)
+ \delta v(s,t) \label{scurv0}
\ee
We get at the linear order
\be
0 = \hat{\Gamma}_0 \delta \boldmath{\mbox{$\phi$}} + \varepsilon {\bf G}(\boldmath{\mbox{$\phi_{\cal D}$}})
+ (\alpha \varepsilon  + \beta k + \delta v) 
\frac{\partial \boldmath{\mbox{$\phi_{\cal D}$}}}{\partial r} + \hat{D} k 
\frac{\partial \boldmath{\mbox{$\phi_{\cal D}$}}}{\partial r} 
+ \varepsilon^{1/2} \boldmath{\mbox{$\Omega$}}(\boldmath{\mbox{$\phi_{\cal D}$}}; r,s;t) 
\label{deltafi0}
\ee
where
\be
\hat{\Gamma}_0 = \hat{D} \frac{\partial^2}{\partial r^2} 
+ v_0
\frac{\partial}{\partial r} + \left.\frac{\partial {\bf f}}
{\partial \boldmath{\mbox{$\phi$}}} \right|_{ \boldmath{\mbox{$\phi$}} = \boldmath{\mbox{$\phi_{\cal D}$}}} 
\label{gamma0}
\ee

Now the right eigenvector of $\hat{\Gamma}_0$ reads
\be
{\bf u_{(0)}} = \frac{\partial \boldmath{\mbox{$\phi_{\cal D}$}}}{\partial r}
\ee 
which is independent of both $\varepsilon$ and $k$.
Taking (\ref{deltafi0}) and performing the scalar product with the left
eigenvector
${\bf u^{(0)}}$, we get
\be
&&k ({\bf u^{(0)}},\hat{D}{\bf u_{(0)}}) + 
(\alpha \varepsilon + \beta k ) ({\bf u^{(0)}}, {\bf u_{(0)}})
+ \varepsilon ({\bf u^{(0)}}, {\bf G}(\boldmath{\mbox{$\phi_{\cal D}$}}))
\nonumber \\ 
&& \qquad \qquad \qquad \qquad + \varepsilon^{1/2} 
({\bf u^{(0)}},\boldmath{\mbox{$\Omega$}}
(\boldmath{\mbox{$\phi_{\cal D}$}}; r,s;t) )
+ \delta v(s,t) ({\bf u^{(0)}}, {\bf u_{(0)}}) 
= 0
\ee
As curvature and fluctuations are linear perturbations we get
\be
\alpha = - \frac{({\bf u^{(0)}},
{\bf G}(\boldmath{\mbox{$\phi_{\cal D}$}}))}{({\bf u^{(0)}},{\bf u_{(0)}})},
\;\;\;\;\;\;\;\;\;\;\;\beta = - \frac{({\bf u^{(0)}}, \hat{D}{\bf u_{(0)}})} 
{({\bf u^{(0)}},{\bf u_{(0)}})}
\ee
and
\be
\delta v(s,t)= -
\varepsilon^{1/2} 
\frac{({\bf u^{(0)}},\boldmath{\mbox{$\Omega$}}(\boldmath{\mbox{$\phi_{\cal D}$}}; r,s;t) )}
{({\bf u^{(0)}},{\bf u_{(0)}})}.
\ee

The stochastic eikonal equation is now
\be
v_n (s,t)= v_0 + \alpha \varepsilon +
\beta \kappa(s,t) + 
D_f^{1/2}(\varepsilon)\zeta(s,t)
\ee
where the noise $\zeta(s,t)$ is a zero mean Gaussian white process with
\be
\langle \zeta(s,t) \zeta(s^{\prime},t^{\prime}) \rangle =
2 \delta (s-s^{\prime}) \delta(t-t^{\prime}),
\ee
and
\be
D_f(\varepsilon) = 
\varepsilon \frac{\int dr \sum_{i,j} u_i^{(0)} u_{j(0)}
g_i(\boldmath{\mbox{$\phi_{\cal D}$}})g_j(\boldmath{\mbox{$\phi_{\cal D}$}})}{({\bf u^{(0)}},{\bf u_{(0)}})^2}.
\ee

\end{document}